\documentclass[12pt]{iopart}
\usepackage{iopams}
\usepackage{harvard}
\usepackage{graphicx}
\usepackage{bbm}
\usepackage{braket}
\usepackage{calc}

\begin{document}

\title[Avalanche Photo-Detection for High Data Rate Applications]{Avalanche Photo-Detection for High Data Rate Applications}

\author{H B Coldenstrodt-Ronge$^{1,2}$ and C Silberhorn$^{1,*}$}

\address{1 Max-Planck Research Group, Guenther-Scharowskystrasse 1/Bau 24, 91058 Erlangen, Germany}

\address{2 Clarendon Laboratory, University of Oxford, Parks Road, Oxford, OX1 3PU, United Kingdom}

\ead{$^*$csilberhorn@optik.uni-erlangen.de}

\begin{abstract}
Avalanche photo detection is commonly used in applications which require single photon sensitivity. We examine the limits of using avalanche photo diodes (APD) for characterising photon statistics at high data rates. To identify the regime of linear APD operation we employ a ps-pulsed diode laser with variable repetition rates between 0.5MHz and 80MHz. We modify the mean optical power of the coherent pulses by applying different levels of well-calibrated attenuation.
The linearity at high repetition rates is limited by the APD dead time and a non-linear response arises at higher photon-numbers due to multiphoton events. Assuming Poissonian input light statistics we ascertain the effective mean photon-number of the incident light with high accuracy. Time multiplexed detectors (TMD) allow to accomplish photon-number resolution by "photon chopping". This detection setup extends the linear response function to higher photon-numbers and statistical methods may be used to compensate for non-linearity. We investigated this effect, compare it to the single APD case and show the validity of the convolution treatment in the TMD data analysis.
\end{abstract}

\submitto{\JOB}

\noindent{\it Avalanche Photo-detection, Photon-number Resolved Measurements, Quantum Communication \/}

\section{Introduction}
Studying distinct quantum properties of light has initiated a multitude of new developments in quantum optics and fundamental physics. In  recent decades it has actually triggered the 
evolution of quantum technologies and quantum communication, which are based on genuine quantum effects that have no classical counterpart. For such applications the interest in the performance of detectors has  shifted: while for fundamental research the quantum efficiency essentially defines  solely the quality of a measurement setup, the benchmark for detectors in quantum communication systems  also comprises the experimental complexity, detector noise and the maximum speed of possible data rates in single-shot operation \cite{Bienfang2004,Gordon2005,Thew2006,Takesue2006}. 

Two general approaches are currently in use to characterise quantum states in communication systems: single-photon schemes rely on avalanche photo-diodes, while  continuous variable (CV) systems employ homodyne detection. 
 Time-resolved homodyne detection setups utilize conventional photo-diodes  to realize single shot measurements of quadrature uncertainties  \cite{Smithey1993,Raymer1995,Hansen2001,Zavatta2002,Wenger2004a}. Thus, CV communication promises quantum key exchange at high repetition rates \cite{Grosshans2002,Hirano2003,Lance2005,Lorenz2006}; though it requires a more intricate data post-processing with reduced sifted secure bit rates. On the contrary, various techniques have been realised up-to-date to observe single photons: photomultiplier tubes, avalanche photo-diodes (APDs), visible light photon counters (VLPCs) and superconducting sensors, like the superconducting edge sensor or superconducting bolometers \cite{Cabrera1998,Fujiwara2006,Hadfield2005,Kim1999,Rosenberg2005,Somani2001}.
In the visible regime avalanche photo-diodes however combine reasonable quantum efficiencies of about 60\% and comparatively low dark count noise  with an operation at room temperature and commercial availability. Neglecting the difference in quantum efficiency, the APDs seem to gather many advantages, but unfortunately they are binary detectors. Thus they are not capable to distinguish directly between different photon-numbers. 
In order to achieve photon-number resolution with APDs, schemes as ``photon chopping'', e.g., by beam splitters \cite{Paul1996,Kok2001}, or time-multiplexing have been proposed and implemented \cite{Banaszek2003,Fitch2003,Achilles2004}. The basic idea is to split the pulse under investigation into several pulses and to measure them subsequently with APDs. The influence of losses and their treatment in the data analysis have been theoretically and experimentally investigated in previous work \cite{Achilles2004,Achilles2006}. Further distortion of  measurement results arises from photons, which are not separated into different bins and thus get still masked by the non-photon resolving nature of the APD.  For the characterisation of photonic states with one single APD  multi-photon contributions have been assessed by attenuation measurements with variable quantum efficiencies. In principle, this method  allows the complete reconstruction of the photon-number statistics \cite{Wenger2004,Zambra2005}.

In this paper we investigate the limits of characterising photon statistics with commercial APDs at 800nm for moderate to high bit rate applications. In this context, a specific parameter of the APDs is their dead time, which ultimately restricts the speed quantum systems can be driven at. Other schemes have suggested to use active, high speed switches in combination with an APD array to overcome correction which arises from detector dead time \cite{Castelletto2006}. Our attempt is to
keep the complexity of the detection to a minimum by utilising pico-second pulsed light at appropriate repetition rates to lessen
dead time limitations. Furthermore, for pulsed systems  dark count contributions can be largely suppressed by applying a narrow time gating.  This is only limited  by the jitter of the APDs and the duration of the light pulses. The full characterisation of APDs involves their count rate response, dark counts, dead times after a detection event and the detection efficiency. Methods of measuring the quantum efficiency of APDs have been proposed and are well established \cite{KLYSHKO1977a,Klyshko1977,Rarity1987,Penin1991,Ware2004}. However, they do not consider effects arising from the dead times for higher count rates.

 We use only passive optical elements and restrict the number of APDs to two. In all experimental configurations we employed coherent light with different intensities, repetition rates and pulse power. While for cw-light the monitored count rates can only be modified by changing the intensity, pulsed light also allows us to control the time slots of possible detection events. 
For comparison we start our analysis by recording the APD response in dependency on well-calibrated intensities of cw-light. We then perform different measurement sets in the pulsed regime with variable repetition rates to separate dead time effects from the influence of multi-photon contributions of incident light. For photon-number resolved measurements with time-multiplexed detection (TMD) the theoretical treatment of higher number contributions is given in \cite{Achilles2004}. We show experimentally that for higher power levels the inclusion of the described convolution effect is essential for the apt interpretation of measured data. Though appropriate theoretical modeling enables a reliable reconstruction of impinging photon statistics. This complements the experimental TMD detector characterisation and shows in combination with the loss inversion \cite{Achilles2006} the capability of TMDs for quantum communication systems. 

\section{Theory of APD photo-detection}

Due to the charge carrier avalanche in the detection process, APDs can resolve individual photons, but as a drawback they saturate already at the single photon level. This binary response reads: either no detection event (\sc no click\rm), or at least one photon is detected (\sc click\rm).  The corresponding POVM elements are given by $\mathbbm{1}-\ket{0}\bra{0}$ and $\ket{0}\bra{0}$ respectively. As long as the photons arrive at the detector individually and well separated in time,
 the detected counts scale linearly with the number of incident photons as intuitively expected.  Timing control can be ensured in the pulsed regime. The response of a binary detector then depends  only on the probability that at least one photon is present in an incident pulse. This probability is given by $P_1=1-P_0$, where $P_0$ denotes the probability that the input light contains no photon. For a detector with quantum efficiency $\eta_{\rm{APD}}$ and a dark count rate $R_{\rm{dark}}$ we can calculate   an expected count rate of 
\begin{eqnarray}
R_{\rm{cnt}}=f_{\rm{rep}}\eta_{\rm{APD}}\left(1-P_0\right)+R_{\rm{dark}},
\label{equ:cntrategeneral}\end{eqnarray}
if a pulse repetition frequency $f_{\rm{rep}}$ is assumed for the input signal states.

In order to characterise the responding behaviour of the APDs, we have to ensure that we  control precisely the influence of the photon statistics of the input light. We used coherent light which is represented in the photon-number basis as
\begin{eqnarray}
\ket\alpha=\sum_n\rm{exp}\left({-\frac{|\alpha|^2}{2}}\right)\frac{\alpha^n}{\sqrt{n!}}\ket{n},
\label{equ_poisson}\end{eqnarray}
i.e. it obeys a Poissonian photon-number distribution. As usual we model the attenuation $1-\eta^2$ as a beam splitter such that the coherent state  transforms like
\begin{eqnarray}
\hat U_\eta\ket{\alpha} =\ket{\eta\alpha}=\sum_n\rm{exp}\left({-\frac{|\eta\alpha|^2}{2}}\right)\frac{\left(\eta\alpha\right)^n}{\sqrt{n!}}\ket{n}.
\label{equ_attenuation}\end{eqnarray}
The probability for no photons present in a pulse is then given by
\begin{eqnarray}
P_0=\rm{exp}\left({-|\eta|^2|\alpha|^2}\right).
\label{equ_naughtphotons}\end{eqnarray}
Assuming a constant quantum efficiency of the APDs, \eref{equ:cntrategeneral} rewrites to
\begin{eqnarray}
\tilde{R}_{\rm{cnt}}=f_{\rm{rep}}\left(1-\rm{exp}\left({-|\eta|^2|\alpha|^2}\right)\right)+\tilde{R}_{\rm{dark}},
\label{equ_cntrateeffective}\end{eqnarray}
with effective values $\tilde{R}_{\rm{cnt}}$ and $\tilde{R}_{\rm{dark}}$.

If the influences on the count rates are known, a count rate dependent correction factor $C\left(R_{\rm{measured}}\right)$ can be introduced. With knowledge of the correction factor the real count rates $R_{\rm{real}}$ can be determined from the measured count rates $R_{\rm{measured}}$:
\begin{eqnarray}
C\left(R_{\rm{measured}}\right)=\frac{R_{\rm{real}}}{R_{\rm{measured}}}.
\label{equ_generalcorrection}\end{eqnarray}
The designation of this factor depends on the calibration of the real photon-number. We explored different possibilities to retrieve the real input photon-numbers for high count rates including the data sheets provided by the manufacturer for the specific APD modules.

\section{Optical setup}

\begin{figure}[htb]
	\noindent
  \centering\includegraphics[width=\textwidth/2]{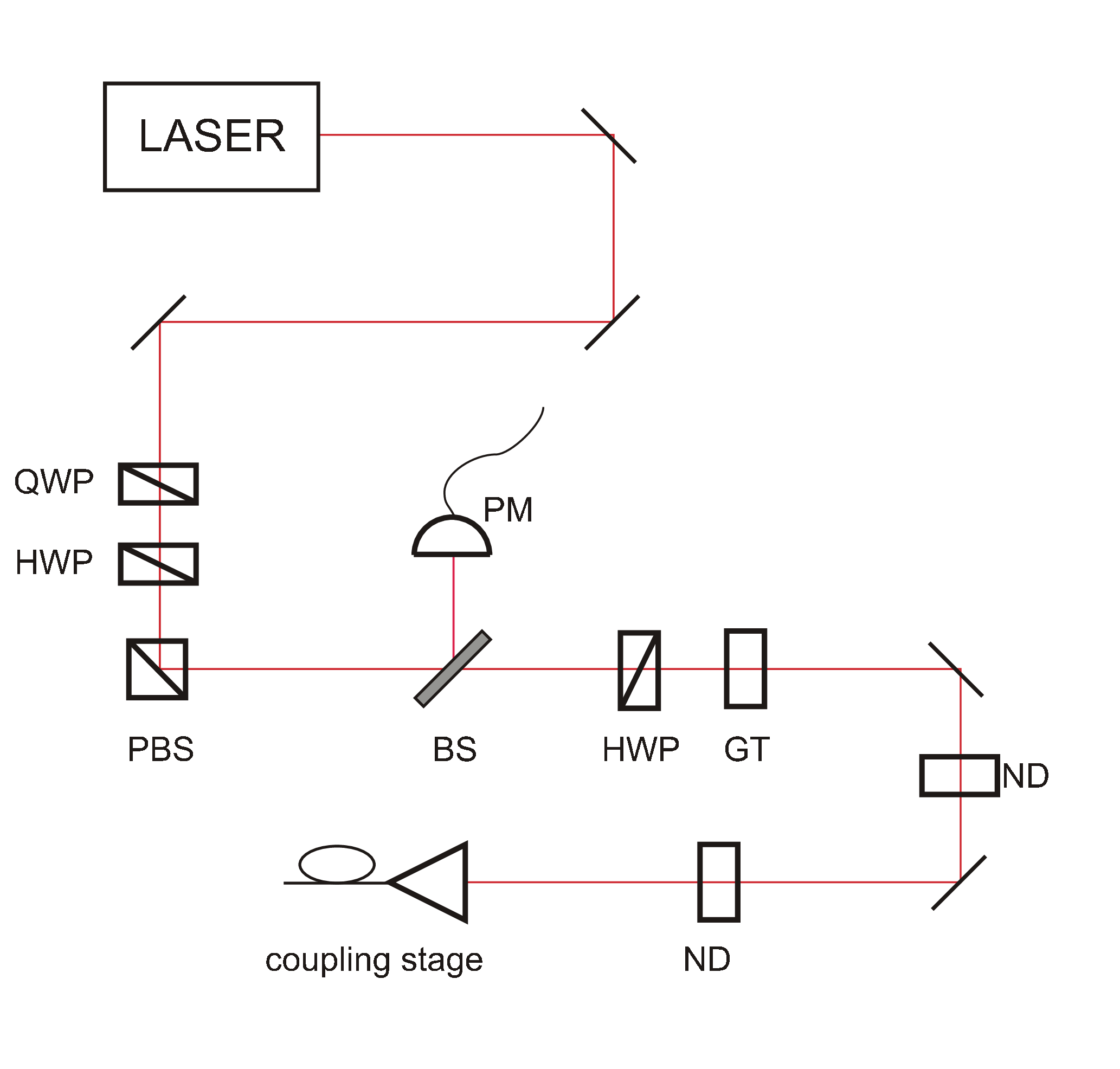}
  \caption{Scheme of the optical setup we used for characterising the APDs in cw- and pulsed operation: QWP: quarter-wave plate, HWP: half-wave plate, PBS: polarising beam splitter, BS: beam sampler, PM: power meter, GT: Glan-Thompson polariser, ND: neutral density filter}
  \label{pic_setup}
\end{figure}

The general experimental setup is illustrated in \fref{pic_setup}. 
 We employed a PicoQuant diode laser system (PDL-800B) for the generation of pulsed coherent states. Our system permitted adjustable repetition rates from less than 1MHz to 80MHz and  delivered ps-pulses  with a centre frequency at $\bar\lambda_{\rm{centre}}\approx$ 805.3nm. The maximum mean power reached up to 1 mW. To realise  the cw-case we used a diode laser with an optical output power of $P_{\rm{cw}}\approx$ 8mW at the centre frequency $\lambda=$ 800nm. For most measurements we utilised a half-wave plate in combination with a polarising beam splitter to implement a variable attenuation  with a dynamical extinction ratio of around 1:35. For the measurements with the cw-laser we used an additional Glan-Thompson polariser, which allowed to access an extinction ratio of more than 1:300.  To ascertain the attenuation factor, part of the light was reflected out  by a beam sampler with  constant splitting ratio and monitored by a 
power meter. In the signal arm a fixed attenuation of several orders of magnitude was necessary to reduce the laser power down to the low photon-number regime. Hereby, special care had to be taken to suppress reflection and interferences between the neutral density filters. These  could otherwise falsify the calibration for different filter sets needed to change the mean optical powers. Furthermore, the power meter used to measure the attenuation showed offsets between its different linearity regimes. Hence all measurements had to be taken with one fixed filter setting and a sufficiently high range of variable attenuation.   Finally, the signal light  was  coupled into a multimode patchcord fibre cable, connected to the APD modules (Perkin\&Elmer SPCM-AQR-13-FC) and a computer based counter card was used to monitor the resulting count rates.

\section{Experimental Results}

 Before recording the data sets for the count rates of various experimental configurations we determined the APD dead times  by taking a histogram of the time difference between two consecutive electronic signals produced by the APDs.
We found a measured APD dead time of $\approx$ 53ns, which was independent of the count rates.
With the APDs in a shielding box, we measured the count rates without any incident light. This absolute dark count level was well below the specified limit of 250 cnt/s.

\subsection{CW-light}

\begin{figure}[htb]
	\noindent
  \includegraphics[width=\textwidth/2]{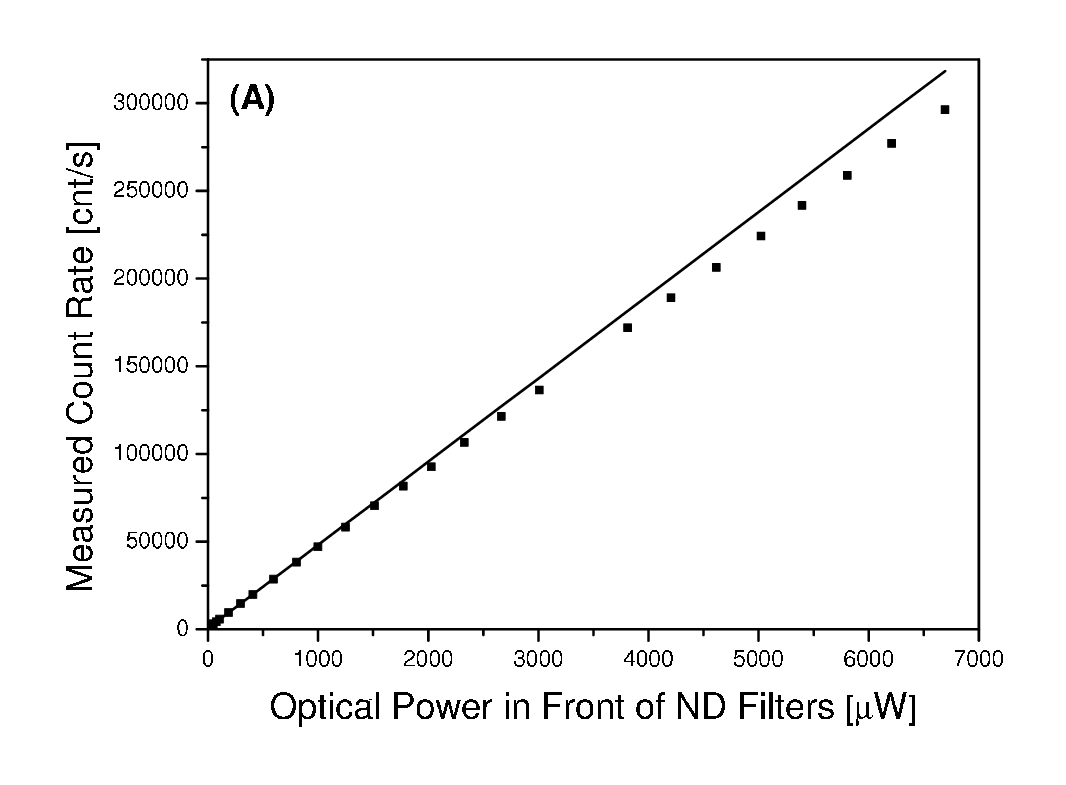}
  \includegraphics[width=\textwidth/2]{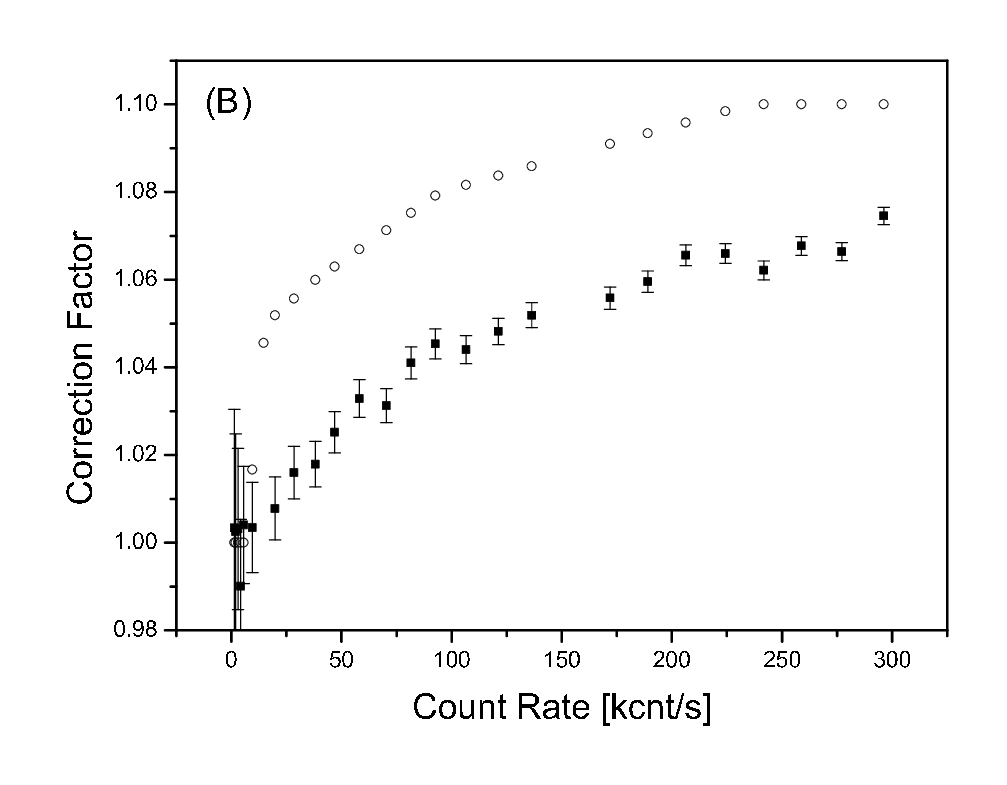}
  \caption{(a) Measured count rate for cw-light against optical power, i.e. transmission of the variable attenuation. The linear fit is included and for higher count rates a significant deviation of the measured data from a linear dependency is visible. The errors in the measured count rates are too small to be visible in the graph. (b) Correction factor calculated according to \eref{equ_generalcorrection} against the measured count rate (\fullsquare) in comparison to interpolated values from the data sheet (\opencircle).}
  \label{pic_cw}
\end{figure}

According to their data sheet \cite{apddata} APDs  show the first signs of a non-linear response at count rates as low as 15.5kcnt/s. Hereby, the manufacturer implicitly assumes cw input light. We recorded APD count rates for cw light at different levels of calibrated attenuation and observed increasing count rates up to 296kcnt/s. In this regime a noticeable correction is already expected. In \fref{pic_cw} we plot the measured count rates against the optical power, i.e. the transmission of the variable attenuation. At low power levels a linear dependency is confirmed such that we can specify a linearity regime for count rates up to 6.5kcnt/s. The corresponding linear fit  is included in the graph.  At higher optical power, the measured count rates are significantly lower than predicted by the fit. The estimated mean photon-number---estimated by  the observed optical powers at the monitoring power meter---is still very low, such that the probability of multi-photon events is negligible at these low count rates.

 The discrepancy of these observed count rates from the linear behaviour can thus not be explained by multi-photon contributions.  If we assume that only the zero photon and single photon components are significant, we can use the linear fit to evaluate the real count rate $R_{\rm{real}}$. According to \eref{equ_generalcorrection} we can now estimate the correction factor $C$ for our experimental data and compare it with the values given in the data sheet of the manufacturer.  \Fref{pic_cw} depicts the corresponding correction factors  in dependence of the detected count rates. Our experimental measurement results show the same general features as the APD data sheet, but they appear to be lower than the specified correction factors of the manufacturer. The instability of the measured values and the global offset are likely to be due to difficulties in the calibration of the {\emph{absolute}} power level for the extremely low light levels we operated at. Note by comparing the equations \eref{equ_poisson} and \eref{equ_attenuation}, that the attenuation of a coherent state yields another coherent state with reduced field amplitude. If the APD detection efficiency is not exactly known, it is only possible to determine the absolute value of field amplitude $\alpha$ of the  coherent input state up to the factor $\eta_{\rm{det}}$. Since the manufacturers only specify a minimum $\eta_{APD}^{min}$ and moreover a precise assessment of the coupling losses is also extremely difficult in this power regime, we have to introduce an effective detection loss parameter $\eta_{\rm{det}}$ to include unspecified coupling and APD efficiencies.

The early non-linearity for the estimated low power level evince that other effects than the high photon-number contributions cause the non-linear behaviour. For cw light the arrival times of the photons are not restricted to specific time slots. After each detection event the APDs are blocked for the duration of their dead time and pulses arriving in this period are not counted. This can happen at any count rate, independently of the higher order photon-numbers and gives rise to the early non-linearity.

\subsection{Pulsed light at constant repetition frequency}

In order to distinguish the impact of the APD dead times on the count rates from the non-linear response caused by the photon-number statistics of the input light, the influence of these two effects needs to be separated in the experiment. The optical power of pulsed light is determined by two degrees of freedom: its pulse energy and its repetition rate. For repetition rates lower than the inverse dead time of the APDs no pulses are  lost during the inactive time of the APDs. 
A repetition frequency of 1MHz provides pulses which are further separated in time than the APD dead time. Thus we expect that the count rates reflect directly the quantum properties of \eref{equ_cntrateeffective} for variable attenuation. 

\begin{figure}[htb]
	\noindent
  \centering\includegraphics[width=\textwidth/2]{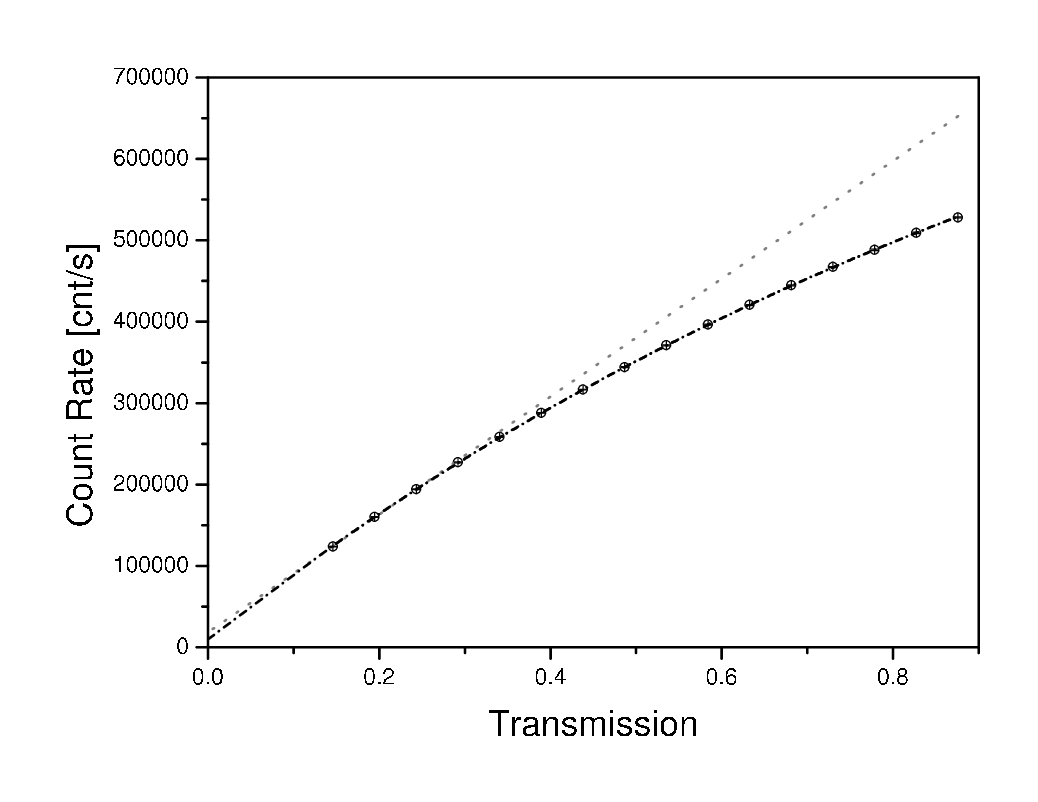}
  \caption{Measured count rates for pulsed light at constant repetition frequency against transmission of the variable attenuator (\opencircle). For low transmission a linear fit (grey, dotted) is still a valid approximation, while for higher transmission the photon-number distribution and binary nature of the detector must be taken into account. A fit following \eref{equ_cntrateeffective} is included (black, dot-line) and shows good agreement with the measured data. The errors in the measured count rates are too small to be visible in the graph.}
  \label{pic:pulsenonlin}
\end{figure}

\Fref{pic:pulsenonlin} shows the detected count rates against the optical transmission with detected values ranging from 124kcnt/s up to 571kcnt/s. For low transmission levels the linear fit included in the figure indicates, that a 
a linear behaviour is again observable. For higher transmission the observed count rates are significantly lower than  predicted by the linear fit. Taking into account the repetition frequency $f_{\rm rep}= 1$ MHz and the photon-number statistics of a coherent state we can estimate the mean photon-number according to \eref{equ_cntrateeffective}. We find an effective mean photon-number of $|\eta_{\rm{det}} \ \alpha|^2$=0.836, with a dark count level of 9.9kcnt/s. While the fitting accuracy of $\pm$0.002 is very good,  the absolute precision of the $|\alpha|^2$ determination suffers again from the missing information of the absolute quantum efficiency of the APDs. 

The fitted dark count level of 9.9kcnt/s is much higher than the absolute dark counts of the APDs, however the fitted dark count levels  are strongly influenced by stray light. These should not be confused with the values obtained for a completely shielded APD during the characterisation of the APDs. The measurement results along with the corresponding fitting curves are shown in \fref{pic:pulsenonlin}. The graphs according to \eref{equ_cntrateeffective}  provides excellent agreement with the measured data. The linear fitting demonstrates that for count rates up to $\approx$230kcnt/s APDs can be linearly approximated. For higher mean photon-numbers the non-photon-number resolving nature of the detection process becomes eminent and the number of counts does not correspond directly to the number of photons any more. Nevertheless the count rates confirm nicely the Poissonian photon-number distribution of coherent light and our modeling of the APD response such that an effective mean photon-number can be retrieved with high accuracy if Poissonian statistics are presumed.

\subsection{Pulsed light at constant pulse energy}

 If the  energy of a single pulse is kept constant, the mean optical power of a pulsed system can also be adjusted  by varying the repetition frequency of the laser. Though, the number of photons per pulse is solely determined by 
the pulse energy. Hence  increasing the mean optical power by going to higher repetition rates corresponds to preparing a larger sample of quantum states with equal properties. In this experimental configuration  non-linear changes in the count rates cannot  are not related to different photon-number properties, but must be attributed to the limited speed of APD detection.

If the repetition time, which is the inverse repetition frequency, is greater than the dead time of the detector, no pulses are masked by a blocked detector. For low pulse energies we expect with \eref{equ:cntrategeneral}  a linear dependency of the count rate on the repetition frequency. For higher repetition frequencies the detector might be in the recovery phase from a previous pulse when a new pulse arrives. This pulse is not counted; we need to introduce a correction for the count rate such that follows
\begin{eqnarray} \label{apd:cnt_rate_pulsed_fr}
R_{\rm{cnt}}=f_{\rm{rep}}\eta_{\rm{APD}}\left(1-P_0\right)C_{\rm{puls}}+R_{\rm{dark}},
\end{eqnarray}
where the additional correction $C_{\rm{puls}}$ is now included. Note, that  $C_{\rm{puls}}$ is not a constant factor, but  depends itself on the repetition rate and the photon statistics.

An event will not be registered if the APD is blocked by a previous detection event. The relevant time scale for the APDs being blocked is their dead time. In order to determine whether a pulse can be detected we need to calculate the probability that no APD counts  occurred prior to a potential detection event during one dead time interval. The repetition rate defines a time grid for possible detection events: for $\frac{T_{\rm{dead}}}{n} \geq T_{\rm{rep}} > \frac{T_{\rm{dead}}}{n+1}$ we find $n$ time slots, which fit into the detector dead time. These must be considered for calculating $C_{\rm{puls}}$. We define $p_\gamma$ as the probability of an APD \sc  click \rm for low repetition rates $p_\gamma=\eta_{\rm{APD}}\left(1-P_0\right)$, which depends also  on the input photon number statistics. The probability of $n$ consecutive time slots being empty ---and thus the detector being ready to fire for a consecutive pulse ---is then given by $\left(1-p_\gamma\right)^n$. This defines the correction factor to first order as

\begin{eqnarray} \label{apd:term_corr_pulsed}
C_{\rm{puls}}=
\cases
{1 & if $T_{\rm{dead}} < \frac{1}{f_{\rm{rep}}}=T_{\rm{rep}}$, \\
1-p_\gamma & if $T_{\rm{dead}} \geq T_{\rm{rep}} > \frac{T_{\rm{dead}}}{2}$, \\
\vdots \\
\left(1-p_\gamma \right)^n & if $ \frac{T_{\rm{dead}}}{n} \geq T_{\rm{rep}} > \frac{T_{\rm{dead}}}{n+1}$. }
\end{eqnarray}
We would like to point out, that this analysis requires low $p_\gamma$.
We neglect effects arising from the possibility that 
pulses can get registered if more than one potential detection event fall within twice the dead time before the considered APD count. For higher $p_\gamma$ such higher order effects must be taken into account and \eref{apd:term_corr_pulsed} overestimates the actually necessary correction.

\begin{figure}[htb]
	\noindent
  \includegraphics[width=\textwidth/2]{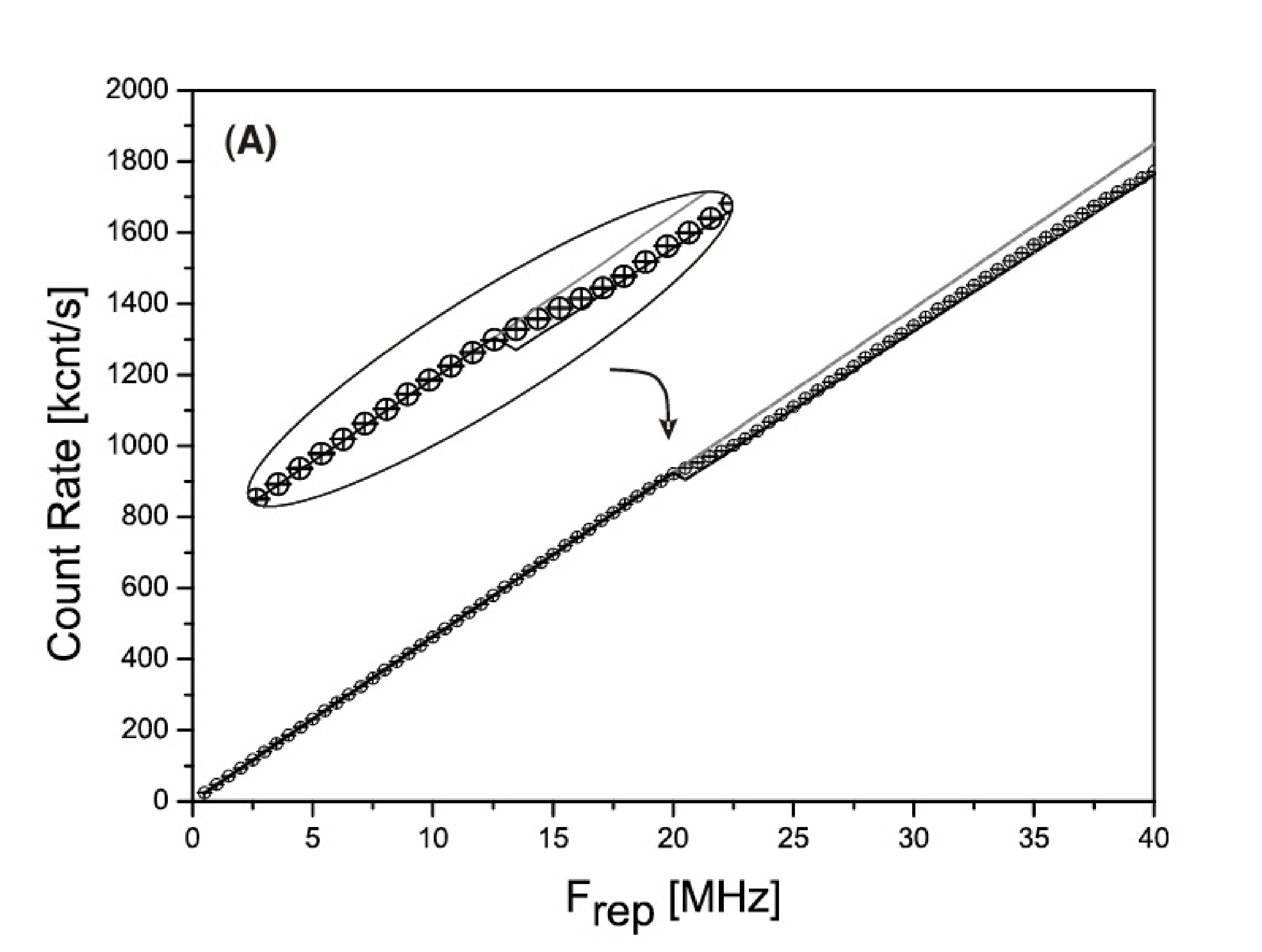}
  \includegraphics[width=\textwidth/2]{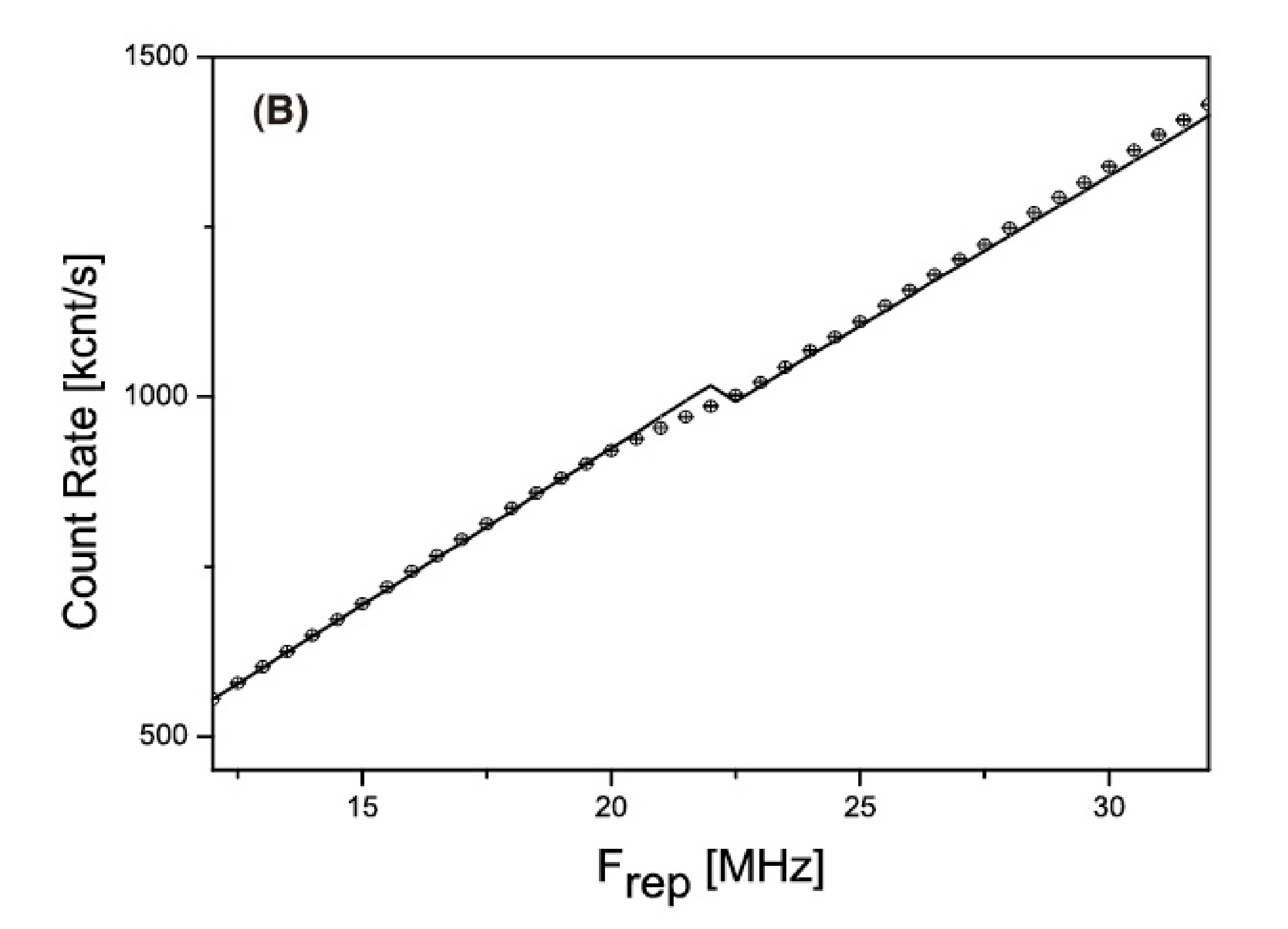}
  \caption{The measured count rates for pulsed light at constant pulse energy are plotted against the laser repetition frequency. (a) The linear fit (light grey) shows the linearity regime up to approximately 15 MHz, while for higher count rates the correction needs to be applied (black line) to explain the non-linear count rate response. (b) The simulated count rate response (black line) reproduces the measured data, if dead time effects are taken into account. For both graphs the errors in the measured count rates are too small to be visible.}
  \label{pic:pulsefrqu}
\end{figure}

For  intervals of the repetition frequency that are defined by  multiples of the inverse dead time, the correction factor does not change.  We expect   linear dependencies within the defined segments. According to \eref{apd:term_corr_pulsed} the different intervals are interrupted by a transition region with a step function. While we were able to reproduce the predicted linearity regimes for repetition frequencies up to 80MHz,  we restrain our plots to the first two linearity regions in order to improve readability.  In \fref{pic:pulsefrqu} we plot the measured count rates against the laser repetition frequency. To cope with power fluctuation we renormalised with respect to the pulse energy at a repetition frequency of 1MHz. Two linearity regimes can be identified in the intervals [0.5MHz;15MHz] and [23MHz;36MHz].

The first linearity regime is valid to repetition rates up to nearly 20MHz, showing count rates of more than 800kcnt/s. 
If we apply the correction factor of the APD data sheet in a naive manner also for pulsed light, it would suggest that 
 significant corrections of up to 10\% are needed to correct for the real count rates. However, the observed linearity thus rules out the necessity and applicability of the standard APD correction to pulsed systems. In contrast, our results show  that for pulsed light we can use APDs without any objections for repetition rate up to their inverse dead times.
In the transition region between the two linearity regimes a drop of count rates can be observed and the transition regime covers the inverse dead time of the detectors. 

We use a binary model for the correction: it takes effect at a certain time and does not take into account dead time jitter. Thus in the transition regime itself the measured data differs from the corrected plot. For the following linearity regime we find good agreement again with the measured data again. We apply our correction factor according to \eref{apd:term_corr_pulsed} to the linear fit, which is included in \fref{pic:pulsefrqu}. The break-in of count rates is clearly accounted for, although the correction slightly underestimated the resulting count rate. After-pulsing and dark count effects scale linearly with the repetition frequency and thus result in an overestimation of $p_\gamma$. Eventually this leads to an overestimate of the necessary correction, which can be understood by recalling 
 that our model is to first order and actually neglects more than two consecutive pulses. As already mentioned it is
 only valid for low probabilities  and relies on the knowledge of the correct value of $p_\gamma$. Still, our results demonstrate clearly the linearity of APDs when operated at frequencies below their inverse dead time  at constant pulse energy and the correction estimated from $p_{\gamma}$ and \eref{apd:term_corr_pulsed} is actually  in good agreement with the experimental data up to a small discrepancy.

\subsection{Simulation}

We designed a Monte-Carlo simulation to validate the predicted influence of the APD dead time. We included cw light sources by modeling the cw count rates as a pulsed system with infinite repetition frequency. A crucial point in this approach is to carefully calculate the pulse energy in order to keep the average optical power constant. For increasing repetition rates the pulse energy must be adjusted using the Poissonian photon-number distribution of coherent light. We simulated the pulsed case first, in order to verify our simulation with the measured data. The simulation reproduced the measured linearity regimes and transition regions. In \fref{pic:pulsefrqu} we compare the measured count rates for a simulation using similar parameters as given in the experiment. Since we used a binary model for the initiation of the correction, the transition region shows a small deviation from the simulation. The linearity regimes are reproduced well by the simulation, but the correction is also overestimated. Since the simulation was initialised with an initial $p_\gamma$ taken from the measurement data this indicates again that afterpulsing and dark count contributions actually caused  the overestimation of this crucial parameter. The significantly lower deviation between simulation and our analytical model arise from some higher order contributions.
Next, we increased the repetition frequency and observed a convergence of the correction factors for high repetition frequencies. These reproduce the shape of the detected correction factor against counts curve in the cw case. An offset is introduced due to insufficient measurement accuracy of the power meter, which is needed for absolute calibration.

The performed experiments in combination with the simulated results provide a complete understanding of the APDs and show the independence of the detection quantum efficiency from the count rate. Corrections only arise from the binary nature of the detector and detector dead time, which might be mistaken as a change in detection efficiency at first glance. Thus for quantum information applications data rates up to the inverse dead time are reasonable.

\section{Application in time multiplexed detection}

TMD measurements rely on the fact that APDs can be driven at rates defined by the inverse of their dead times while showing no drop of their quantum efficiencies.
We have experimentally confirmed  these assumptions with the results presented in the previous sections. The concept of TMDs offers the distinct advantage of low experimental complexity while the speed of operation is only limited by the minimum time APDs require to measure consecutive detection events.

In TMD detection the photon-number resolution is achieved by using temporal modes for "photon chopping":  a pulse signal which is under investigation is split into several pulses, which get partially delayed in time and detected with individual binary detectors. For the practical realisation of a TMD  an important  issue is the setting  of the base delay between expected detection events. This must be at least the APD dead time in order to prevent the temporal modes from being masked by the APD dead time.
To test the applicability of TMDs for higher bit rate applications we implemented a two-stage TMD with eight outgoing modes.  We actually chose a delay around $\Delta T_{\rm{base}}$=100ns to restrain also after-pulsing influences. Additionally to previous realisations, we implemented an electronic gating to inhibit noise contributions and a computer controlled data acquisition system. The data acquisition speed above 2 MHz in our system is only bounded by physical constraints, i.e. four times the base delay.

For light with non-negligible higher multi-photon contributions  multiple photons may still be transmitted into the same mode when leaving the TMD fiber network. They get not discriminated by the subsequent detection with the binary APDs, resulting in a lower count rate than the actual photon number of the input light. While this introduces an uncertainty for one single-shot measurements,  for ensemble measurements this experimental imperfection can be included in the post-processing of the data analysis for studying photon-number statistics. As shown previously  \cite{Achilles2004} a photon-number distribution $\vec{\rho}$ transforms to click statistics $\vec{p}$ according to 
\begin{eqnarray}
\vec{p}=\mathbf{C}\vec{\rho},
\end{eqnarray}
where the convolution matrix $\mathbf{C}$  accounts for the probabilities of $n$ incident photons resulting in $0\ldots n$ detectors firing. 

Losses in the optical system and the non-unity detector efficiency also result in lower click rates than photons in the initial pulse. All losses in the system can be combined in an effective single beam splitter in front of the fibre network \cite{Silberhorn2004,Achilles2006} and represented by an additional matrix $\mathbf{L}$. The initial photon-number distribution is then retrieved from the detector click statistics using an ordinary matrix inversion:
\begin{eqnarray}
\vec{\rho}=\mathbf{L}^{-1}\mathbf{C}^{-1}\vec{p}.
\end{eqnarray}

While the effects of losses and the usage of the TMD detection for the characterization of arbitrary photon statistics are thoroughly investigated in previous work \cite{Achilles2006,Achilles2004} we concentrate on the influence arising due to the convolution. We utilised an optical setup similar to the experiments with APDs shown in \fref{pic_setup} to feed coherent light of different power levels into the TMD. We would like to point out that that the usage of coherent light in these studies is crucial  in order to be able to distinguish effects arising from attenuating the input state statistics from distinct characteristics, which are intrinsic to the TMD detector response. As stated earlier, attenuation only changes the mean photon-number of the Poissonian statistics while effects on the shape of the photon number distribution caused by  introducing losses are eliminated. The mean number of photons depends linearly on the attenuation. Though in analogy to a single APD being illuminated with more than a single photon per pulse, masking of higher photon-numbers due to photons, which stay together in one outgoing mode, is expected. 

\begin{figure}[htb]
	\noindent
  \includegraphics[width=\textwidth/2]{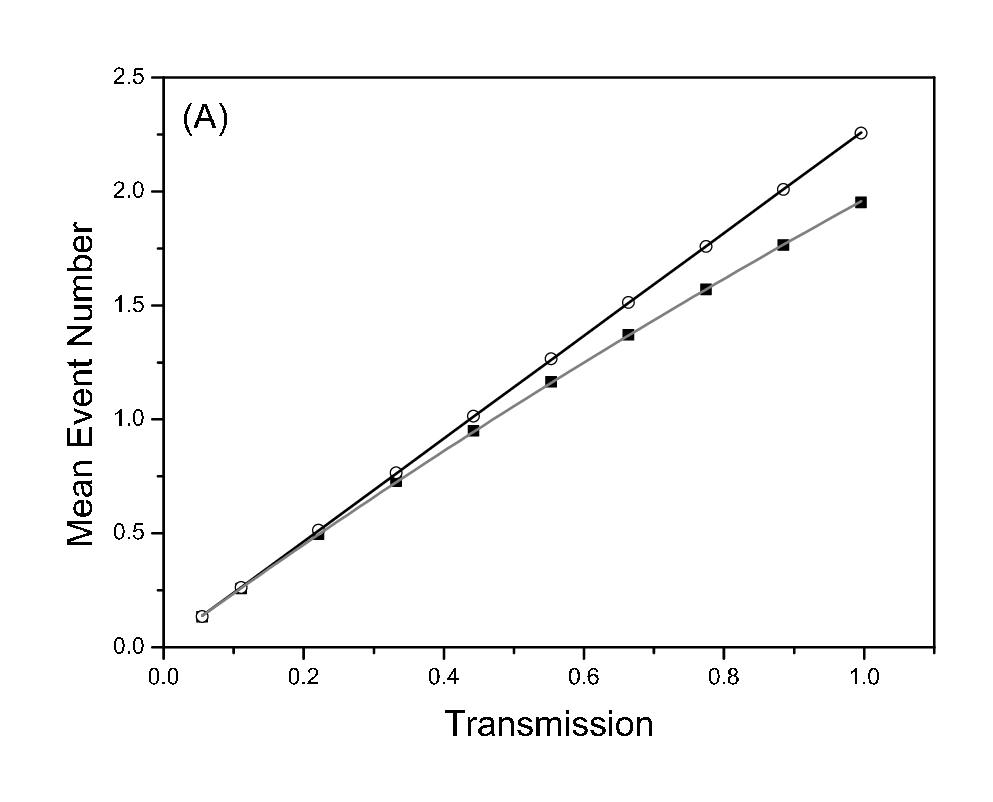}
  \includegraphics[width=\textwidth/2]{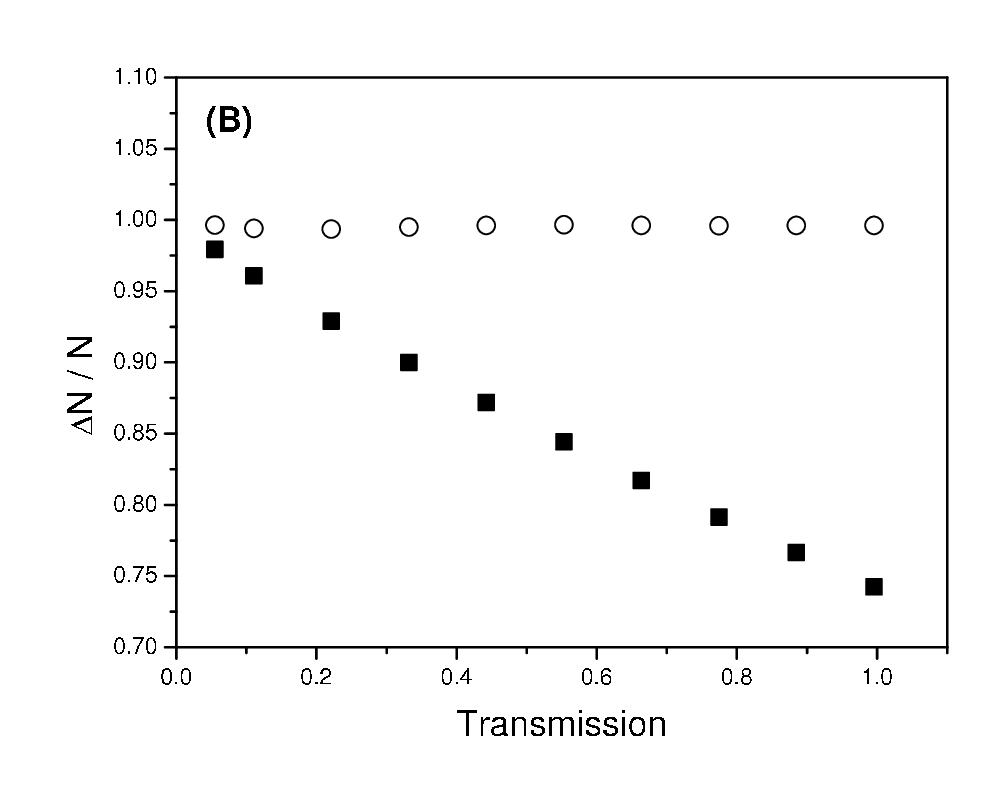}
  \caption{(a) The mean number of clicks (\fullsquare) and mean number of photons after deconvolution (\opencircle) are plotted against the transmission. For the deconvoluted data the expected linear fit holds over the full range of transmission, while for the click data non-linear behaviour is visible for higher transmission. The clicks follow \eref{TMD:theorieverlust2}, which can be verified looking at the light grey fitting line. (b) For the deconvoluted data (\opencircle) the Mandel Q-parameter against the transmission shows the expected value of 1 for all transmission levels. For the click data Mandel Q-parameters less than 1 falsely indicate non-Poissonian nature.}
  \label{pic:meannumbers}
\end{figure}

In \fref{pic:meannumbers} we plot the mean number of clicks before and photons after deconvolution against the applied attenuation. Similar to the result of \fref{pic:pulsenonlin} a linear dependency is visible for low transmissions, corresponding to a low mean number of clicks or photons respectively. For higher click rates a significant deviation from the linear fit develops, while the deconvoluted data increases linearly with increasing transmission. More photons are distributed into the same time slot, triggering less clicks than photons in the pulse. Thus the raw data will give an underestimate of the mean photon-number, which can be modeled by assuming $k$ perfect beam splitters in the fibre network. The coherent input state with mean photon-number $|\alpha|^2$ is split into $2^k$ modes with corresponding mean photon-numbers ${|\alpha|^2}/{2^k}$. For each mode the detection rate is given by \eref{equ_cntrateeffective} and summing these rates yields the effective count rate after the fibre network as
\begin{eqnarray}
\tilde{R}_{\rm{cnt}}=\sum^{2^k}_{i=1} f_{\rm{rep}}\left(1-\rm{exp}\left({-|\eta|^2\frac{|\alpha|^2}{2^k}}\right)\right)+\tilde{R}_{\rm{dark}}.
\label{TMD:theorieverlust}\end{eqnarray} 
The expected mean number of clicks is obtained by dividing by the repetition frequency and for a TMD with eight time bins  we expect a mean number of clicks
\begin{eqnarray}
\bar{c}=8\left(1-\rm{exp}\left({-|\eta|^2\frac{|\alpha|^2}{8}}\right)\right)+\frac{\tilde{R}_{\rm{dark}}}{f_{\rm{rep}}}.
\label{TMD:theorieverlust2}\end{eqnarray}
In \fref{pic:meannumbers}(A) we also display a fit of this expected click rate to the data, which shows perfect correspondence for values of $\tilde{R}_{\rm{dark}}$=16kcnt/s$\pm$3kcnt/s and an effective $|\alpha|^2$=2.232$\pm$0.006. The graphs indicate that we are able to demonstrate experimentally convolution utilizing well known quantum states. Moreover, the use of the deconvolution method to retrieve the original statistics is experimentally verified by retrieving the linear dependence.

The value of the Mandel Q-parameter ${\sigma^2}/{\bar{n}}$ is often used to characterise the non-classicality of a photon-number distribution\cite{Mandel1995}. In \fref{pic:meannumbers} we plot the Q-parameter against the variable transmission and compare the results for the raw click statistics with the deconvoluted data.
A Poissonian distribution of a coherent state always results in a Q-factor of 1. Our experimental data confirm this expected Q-factor for the deconvoluted data at all transmission levels. For low photon-numbers, i.e. transmission, the raw data still show values  near 1.  This value, however,  drops  significantly below 1 for higher transmission rates,  indicating falsely a non-Poissonian nature of the assumed photon-number statistics. Higher photon-numbers are more strongly affected by the decrease in clicks than lower photon-numbers. Thus the raw click statistics show non-Poissonian features indeed, which must not be mistaken for non-Poissonian photon-number statistics. From our measurements---depicted in \fref{pic:meannumbers}(B)---we ascertain that the deconvolution of the measured click statistics is essential to obtain correct photon-number statistics when utilizing TMD detection for quantum state characterisation. Contrariwise, we demonstrated experimentally that using the deconvolution matrix approach the correct statistics can be deduced.

Since the Q-parameter is regarded as an indicator for the non-classicality of light, it is crucial to consider the impact of detectors on measured outcomes. At first glance---without taking detector saturation into account---a source might appear to exhibit non-classical Q-parameters which might then be falsely interpreted as photon-number squeezing. In our parameter regime  we demonstrated that the TMD detector saturation can be perfectly compensated by applying the deconvolution in the data post-processing. 
This is verified by the constant Q-factor of 1 after the application of the deconvolution. 
The  computation power for calculating the convolution matrix numerically rises exponentially with the number of photons resolved by the detector. For more than eight TMD bins only the idealised case of perfect splitting ratios can be easily calculated analytically \cite{Fitch2003}. Thus we investigated the effects of using symmetric splitting ratios for  deconvoluting simulated data with unbalanced splitting ratios. We  found that the convolution matrices are surprisingly independent of the splitting ratios of the beam splitters used in the TMD fibre network -- which, in turn, enables the extension of the TMD to higher numbers of bins without any major problems in the mathematical treatment. In practice 
the deviations due to convolution effects will be much less than errors caused by imperfections in the optical loss calibration.

\section{Conclusion and Discussion}

In summary we characterised APDs in the context of testing the limits of their application in quantum information processing. A major challenge is the extremely precise calibration of the used attenuation, which contributes most to the errors of the experiment. In particular the cw-light measurements require optical power spanning several orders of magnitude in order to start measuring in the linear regime, while getting significant non-linearities at higher power levels.
With the simulation being extremely sensitive to errors in the calibration, better methods for even more precise calibration are needed and should close the gap between simulation and measured correction factors.

By using pulsed light we could differentiate two sources of non-linear APD response: intensity and pulse repetition frequency. For quantum  communication at high data rates an essential characteristic of an APD is its dead time where we verified for repetition frequencies up to its inverse dead time a linear APD behaviour. With dead times of 50ns for commercially available APD modules this allows for data rates up to 20MHz.

An important application of APDs is the TMD, where we complemented the detector description by verifying experimentally that saturation effects can be compensated in the post-processing data analysis. We can now operate the TMD at the data acquisition speed limit given by the APDs dead time. This is a powerful tool to quickly characterise quantum states in the photon-number bases allowing the collection of large number of statistical data in short measurement time. We expect that this provides the basis for photon-number based characterisation of highly non-classical quantum states where the number of recorded data points during the stability of an optical setup becomes crucial.

\section*{Acknowledgment}
This work has been supported by the European Commission under the Integrated Project Qubit Applications (QAP) funded by the IST directorate as Contract Number 015848.

\end{document}